\pdfoutput=1
\documentclass[usenatbib,useAMS]{mn2e}
\usepackage{times}
\usepackage{snapshot}
\usepackage[a4paper, totalwidth=480pt, totalheight=680pt]{geometry}
\usepackage{graphicx}
\usepackage{epstopdf}
\newcommand{\kms}{\mbox{km s$^{-1}$}}   %\kms symbol
\newcommand{\ms}{\mbox{m s$^{-1}$}}
\usepackage{aecompl}

\title[Rossiter-McLaughlin observations of WASP-13b and WASP-32b]
{A Window on Exoplanet Dynamical Histories: Rossiter-McLaughlin observations of WASP-13b and WASP-32b}

\author[R. D.~Brothwell et al.]
{R. D.~Brothwell$^1$\thanks{Email: rbrothwell01@qub.ac.uk}, C. A.~Watson$^1$, G.~H\'{e}brard$^{2,3}$, A. H. M. J. Triaud$^{4,5}$, H. M.~ Cegla$^1$, 
\newauthor A. Santerne$^{6}$, E. H\'{e}brard$^{7}$, D. R. Anderson$^{8}$, D.~Pollacco$^{9}$, E. K.~Simpson$^1$, F.~Bouchy$^{2,3}$, 
\newauthor D. J. A. Brown$^{9}$, Y. G\'omez Maqueo Chew$^{9}$, A.~Collier Cameron$^{10}$, D. J. Armstrong$^{9}$,
\newauthor S. C. C.~Barros$^{11}$,  J. Bento$^{9,12}$,  J. Bochinski$^{13}$,  V. Burwitz$^{14}$, R. Busuttil$^{13}$, L. Delrez$^{15}$, 
\newauthor A. P.\ Doyle$^{8}$, F. Faedi$^{9}$, A. Fumel$^{15}$, M. Gillon$^{15}$, C. A. Haswell$^{13}$, C. Hellier$^{8}$, 
\newauthor E. Jehin$^{15}$, U. Kolb$^{13}$, M. Lendl$^{4}$, C. Liebig$^{10}$, P. F. L. Maxted$^{8}$, J. McCormac$^{16,9}$,
\newauthor G. R. M. Miller$^{10}$, A. J. Norton$^{13}$, F. Pepe$^{4}$, D. Queloz$^{17}$, J. Rodr\'iguez$^{18}$, D. S\'egransan$^{4}$, 
\newauthor I. Skillen$^{16}$, B. Smalley$^{8}$, K. G. Stassun$^{19,20}$, S. Udry$^{4}$, R. G. West$^{9}$, P. J. Wheatley$^{9}$ \\
$^1$ Astrophysics Research Centre, School of Mathematics \& Physics, Queen's University Belfast, BT7 1NN, UK \\
$^2$ Institut d'Astrophysique de Paris, UMR7095 CNRS, Universit\'e Pierre \& Marie Curie, France \\
$^3$ Observatoire de Haute-Provence, CNRS/OAMP, 04870 St Michel l'Observatoire, France \\
$^4$ Observatoire astronomique de l'Universit\'e de Gen\`eve, 51 ch.\ des Maillettes, 1290 Sauverny, Switzerland \\
$^5$ Department of Physics, and Kavli Institute for Astrophysics and Space Research, Massachusetts Institute of Technology, Cambridge, MA 02139, USA \\
$^6$ Centro de Astrofisica, Universidade do Porto, Rua das Estrellas, 4150-762, Porto, Portugal \\
$^7$ IRAP-UMR 5277, CNRS and Univ. de Toulouse, 14 Av. E. Belin, F-31400, Toulouse, France \\
$^{8}$ Astrophysics Group, Keele University,  Staffordshire, ST5 5BG, UK\\
$^{9}$ Department of Physics, University of Warwick, Coventry, CV4 7AL \\
$^{10}$ School of Physics and Astronomy, University of St Andrews, North Haugh, St Andrews, Fife KY16 9SS, UK\\
$^{11}$ Aix Marseille Universit\'e, CNRS, LAM (Laboratoire d'Astrophysique de Marseille) UMR 7326, 13388, Marseille, France \\ 
$^{12}$ Department of Physics and Astronomy, Macquarie University, NSW 2109, Australia \\
$^{13}$ Department of Physical Sciences, The Open University, Milton Keynes, MK7 6AA, UK \\
$^{14}$ Max Planck Institut f\"ur Extraterrestrische Physik, Giessenbachstrasse 1, 85748 Garching, Germany \\
$^{15}$ Universit\'e de Li\`ege, All\'ee du 6 ao$\hat {\rm u}$t 17, Sart Tilman, Li\`ege 1, Belgium \\
$^{16}$ Isaac Newton Group of Telescopes, Apartado de Correos 321, E-38700 Santa Cruz de Palma, Spain  \\ 
$^{17}$ Department of Physics, University of Cambridge, J J Thomson Av, Cambridge, CB3 0HE, UK \\
$^{18}$ Observatori Astron\`omic de Mallorca, Cam\'i de l'Observatori s/n 07144 Costitx, Mallorca, Spain \\  
$^{19}$ Physics and Astronomy Department, Vanderbilt University, Nashville, Tennessee 37235, USA \\
$^{20}$ Department of Physics, Fisk University, Nashville, Tennessee 37208, USA\\}

\date{Accepted 2014 March 12.  Received 2014 March 12; in original form 2013 November 25}

\begin{document}
\raggedbottom

\maketitle

\begin{abstract}
\noindent
We present Rossiter-McLaughlin observations of WASP-13b and WASP-32b and determine the sky-projected angle between the normal of the planetary orbit and the stellar rotation axis ($\lambda$). WASP-13b and WASP-32b both have prograde orbits and are consistent with alignment with measured sky-projected angles of $\lambda={8^{\circ}}^{+13}_{-12}$ and $\lambda={-2^{\circ}}^{+17}_{-19}$, respectively. 
 
Both WASP-13 and WASP-32 have $T_{\mathrm{eff}}<6250$K and therefore these systems support the general trend that aligned planetary systems are preferentially found orbiting cool host stars. A Lomb-Scargle periodogram analysis was carried out on archival SuperWASP data for both systems. A statistically significant stellar rotation period detection (above 99.9\% confidence) was identified for the WASP-32 system with $P_{\mathrm{rot}}=11.6 \pm 1.0 $ days. This rotation period is in agreement with the predicted stellar rotation period calculated from the stellar radius, $R_{\star}$, and $v \sin i$ if a stellar inclination of $i_{\star}=90^{\circ}$ is assumed. With the determined rotation period, the true 3D angle between the stellar rotation axis and the planetary orbit, $\psi$, was found to be $\psi=11^{\circ} \pm 14$. We conclude with a discussion on the alignment of systems around cool host stars with $T_{\mathrm{eff}}<6150$K by calculating the tidal dissipation timescale. We find that systems with short tidal dissipation timescales are preferentially aligned and systems with long tidal dissipation timescales have a broad range of obliquities.
\end{abstract}


\section{Introduction}\label{Intro}

The study of gas giants orbiting close to their host stars allows an insight into the formation and evolution of exoplanets. For example, combined planetary transit photometry and radial velocity (RV) measurements enables the planetary density to be found, providing constraints on the planetary composition. Whilst this provides clues to the formation processes at work, the Rossiter-McLaughlin (RM) effect is thought to be a complementary probe of exoplanet dynamical histories. The RM effect is measured using in-transit spectroscopic observations, revealing a deviation from the Keplerian orbital motion as the star orbits the barycentre of the star-planet system. The effect is caused by the planet occulting the rotating stellar surface. This introduces an asymmetry in the  stellar absorption profile, resulting in an apparent shift of the spectral lines. The RM waveform allows the sky projected spin-orbit alignment angle ($\lambda$) between the rotation axis of the host star and the normal to the planetary orbital plane to be determined. 
       
The alignment angle is thought to provide a window on the dynamical evolution of exoplanets. Hot-Jupiters are thought to form beyond the snow-line where icy cores become massive enough to accrete a gaseous envelope before subsequently migrating either via planet-disk, planet-planet or planet-star interactions. Planet-disk interactions are thought to be dynamically gentle \citep{Goldreich1980, Lin1996} and do not peturb the original inclination of the planet. Other migration mechanisms such as planet-planet and planet-star interactions via the Kozai-Lidov mechanism are more dynamically violent \citep{Kozai1962, Lidov1962}. The presence of a third body in the system excites periodic oscillations in the eccentricity and inclination of the orbit, where tidal dissipation and circularisation shrinks the semi-major axis. The oscillating inclination resulting from Kozai-Lidov interactions produces a continuum of inclinations with stable orbits. Thus, it is expected that hot-Jupiters will exhibit misaligned orbits if such migration processes are operating.

However, it should be noted that measuring a spin-orbit alignment angle of $\lambda=0^{\circ}$ does not necessarily indicate an aligned planetary system. When the impact parameter is low, the RM waveform is independent of $\lambda$ and instead controls the amplitude, leading to a strong degeneracy between $v \sin i$ and $\lambda$ \citep{Gaudi2007}. For example, in a system with an impact parameter of $b=0$ and/or where the stellar rotation axis is inclined in the direction of the observer, any orientation of the planetary orbit leads to a symmetric RM waveform. By calculating the inclination of the stellar rotation axis, these degeneracies can be broken and the true `3D' system geometry ascertained.    

Currently, 76\footnote{Holt-Rossiter-McLaughlin Encyclopaedia: \newline http://www.physics.mcmaster.ca/~rheller/} planets have a measured $\lambda$ where 45$\%$ of planets show substantial misalignments. This population of misaligned planets appears to be synonymous with hotter host stars ($T_\mathrm{eff} \geq 6250$K) whilst aligned planets are preferentially observed orbiting cool host stars. One proposed reason for the alignment-misalignment transition is a change in the internal structure of main-sequence host stars around 6250K, where the outer convective envelope is responsible for tidal interactions. Another correlation in current RM data is the degree of alignment with system age \citep{Triaud2011angle}. All systems with M$_{\star} \geq 1.2$ M$_{\odot}$ were considered and systems with ages greater than 2.5 Gyrs are preferentially aligned, whereas those below this age are preferentially misaligned. This reflects the development of the convective envelope with system age and lends further support to alignment arising from tidal interactions. \citet{Albrecht2012} showed that other correlations of alignment with the orbital period, ratio of planet mass to stellar mass and possibly orbital distance with $\lambda$ provide further evidence that realignment is driven by tidal interactions.   

In order to interpret the results of RM observations as a tracer of dynamical evolution alone, it must be assumed that the original protoplanetary disk and the star are well-aligned. While this seems valid based on angular momentum conservation, theoretical models have begun to challenge this assumption, showing that star-disk misalignment is possible in the pre-mainsequence phase \citep{Bate2010, Lai2011}. Thus, measuring $\lambda$ may not trace planet migration mechanisms but perhaps traces star formation processes or a combination of both. \citet{Watson2011} studied the inclination of resolved debris disks and the inclination of their host stars for 9 systems, showing that all are consistent with alignment. The authors note that all systems have $T_\mathrm{eff} < 6250$K and other candidates with $T_\mathrm{eff} > 6250$K would be important in exploring the full alignment-misalignment theoretical picture proposed by \citet{Winn2010}. Further systems, with a range of spectral types, were investigated by \citet{Greaves2013} where the stellar inclination was found to be aligned with the spatially resolved debris disk for all systems. Recently \citet{Kennedy2013} tested the alignment of the full star-disk-planet system for HD 82943, the first time the full alignment of a system has been investigated. The complete system (the inclination of the stellar rotation axis, the normal to the disk plane and the normal to the planetary orbit) was found to be aligned at a level similar to the Solar System. 

Another approach to distinguish between primordial star-disk misalignments and misalignment driven by migration is to consider the growing number of multiplanet systems. \citet{Albrecht2013} recently analysed the multiple-transiting systems KOI-94 \citep{Hirano2012} and Kepler-25, finding $\lambda=-11 \pm 11^{\circ}$ and $\lambda=7 \pm 8^{\circ}$, results consistent with alignment. Whilst this was thought to hint that multi-planet systems migrate via planet-disk interactions and hot-Jupiters migrate by a different pathway, evidence for misaligned multi-planet systems has been found \citep{Huber2013, Walkowicz2013}. It is clear that a full picture of hot-Jupiter formation and migration is far from complete, requiring the continual buidling of statistics, preferably beyond the $T_\mathrm{eff}$ dependence, to explore unstudied regions of parameter space.

In this paper we report Rossiter-McLaughlin (RM) observations of WASP-13 and WASP-32. WASP-13 and WASP-32 are both slow rotators \citep{Skillen2009, Maxted2010} and cool stars with effective temperatures $\sim$6000K. Section 2 outlines the observations and analysis procedure. In Section 3 the derived parameters are presented and discussed. Next a search for the stellar rotation period for both systems was investigated. For WASP-32, where a period was found, we then computed the true 3D alignment angle. Finally, we conclude with a discussion of our results in Section 4.

\begin{table*}
	\centering
	\caption{Adopted system parameters and uncertainties used to model the RM effect, and other photometric parameters used in this work. The reference is indicated at the end of the column for each object.}	
	\label{Adopted}
	\begin{tabular}{lccc}
	\hline\hline
	Parameter (units)			&Symbol & {WASP-13}  & {WASP-32} \\
	\hline
	Orbital Period (days) 				&$P$   &$~4.3530135\pm~0.0000027$	  &$~2.718659\pm~0.000008$ \\[4pt]
	
	Transit epoch	&$T_\mathrm{0}$		&$2455305.62823 \pm~0.00025~(\mathrm{BJD_{UTC}})$ &  $2455151.0546\pm~0.0005~(\mathrm{HJD})$ \\[4pt]
	
	Transit duration (hours)		&$T_{d}$				&$4.003\,\pm~0.024$	&$2.424\pm~0.048$ \\[4pt]
	
	Orbital inclination ($^{\circ}$)		& $i$ 				&$85.43\,\pm~0.29$	&$85.3\,\pm~0.5$ \\[4pt]
	
	Planet/Star radius ratio		& $R_{p}/R_{*}$		&$0.0919\pm~0.0126$	  &$0.11\,\pm~0.01$ 	\\[4pt]
   
   	Scaled semi-major axis		&$a/R_{*}$ 			&$7.54\pm~0.27$	   &$7.63\,\pm~0.35$ \\[4pt]

	Eccentricity			   &$e$				&$0\,\mathrm{(adopted)}$		&$0.0180\pm~0.0065$ \\[4pt]

	\hline

Reference		& & \citet{Chew2013}  & \citet{Maxted2010} \\

	\hline

	\end{tabular}
	\end{table*}

\section{Data Analysis} \label{OandM}
 
\subsection{Observations and data reduction}

All in-transit radial velocity (RV) data for WASP-13 and WASP-32 were obtained using the SOPHIE spectrograph mounted on the 1.93m telescope at the Observatoire Haute Provence (OHP). SOPHIE is an environmentally stabilised echelle spectrograph (wavelength range 382-693nm) designed for high-precision RV measurements. Two 3 arcsecond optical fibres were used, with one centred on the target and the other used to simultaneously monitor the sky background in case of lunar light contamination. The spectra were then reduced using the SOPHIE data reduction pipeline (Perruchot et al. 2008). Radial velocities were extracted using a weighted cross-correlation of each spectrum with a G2 spectral-type mask. A Gaussian was then fitted to the resulting cross-correlation functions to obtain the radial velocity shift. Uncertainities were computed using the empirical relation of Bouchy et al. (2009) and Cameron et al. (2007). The observation and data reduction details particular to each system are presented in Section 3. 

\subsection{Determination of the system parameters}\label{sec:params}
	
The RM effect and orbit were fitted simultaneously using all the available spectroscopic data, including previously published orbital data. A Keplerian model was used for the orbit, and the analytical approach described in \citealt{Ohta2005} (hereafter OTS) was used to model the RM effect.  An independent systemic velocity was fitted to each orbital dataset in order to account for any instrumental offsets. Similarly the transit datasets were fitted with separate systemic velocities to incorporate instrumental and long-term stellar activity variations.

To fit the RM effect the OTS equations were modified to make them dependent on $R_{p}/R_{*}$ and $a/R_{*}$ rather than $R_{p}$, $R_{*}$ and $a$, to reflect the parameters derived from photometry, and reduce the number of free parameters. The model comprises 11 parameters: the orbital period, $P$; mid-transit time, $T_\mathrm{0}$; planetary to stellar radius ratio, $R_{p}/R_{*}$; scaled semi-major axis, $a/R_*$;  orbital inclination, $i$; orbital eccentricity, $e$; longitude of periastron, $\omega$; radial velocity semi-amplitude of the host star, $K$; sky projected angle between the stellar rotation axis and orbital angular momentum vector, $\lambda$; projected stellar rotational velocity, $v \sin {i} $ and the stellar linear limb-darkening coefficient, $u$.

In summary, the OTS model assumes that the star and transiting planet are disks where the planet is an opaque occulting disk. The radial velocity of a small element on the stellar disk is given by multiplying the $x$-position of the element (Figure 3 of OTS) by $v \sin i$. This quantity is then weighted by the intensity of the stellar disc at that location and then all the elements are integrated over the entire stellar surface. The OTS equations (see sections 5.1 and 5.2 of OTS) result from assuming a linear limb darkening law for the stellar intensity. A linear limb darkening law is assumed as the quadratic model is known to deviate by only a few \ms from the linear limb darkening model. Also, it has been shown that by setting $u$ as a free parameter, $\lambda$ and $v \sin i$ are not significantly affected \citep{Simpson2011a}.   

A series of parameters included in the model have been derived previously from transit observations ($P$, $R_{p}/R_{*}$, $a/R_{*}$ and $i_{p}$). These constraints can be included in the fit in the form of a $\chi ^{2}$ penalty function:

\begin{eqnarray} \label{eqw1}
& \chi^2 & =  \sum_{i}^{} \left [ \frac{v_{i, \mathrm{obs}} - v_{i, \mathrm{calc}}}{\sigma_{i}} \right ]^{2}  + \;\;\;\nonumber \\
&  &     \left(  \frac{X - X_{\mathrm{obs}}  + [\sigma _{X_{\mathrm{obs}}} \times G(0,1)] }{\sigma _{X_{\mathrm{obs}}}} \right)  ^{2} \;\;\;\nonumber \\
\end{eqnarray}
\noindent
where $v_{i, \mathrm{obs}}$ and $v_{i, \mathrm{calc}}$ are the $i$th observed and calculated radial velocities from the model, respectively, and $\sigma_{i}$ is the corresponding observational error. $X$ is one of the fitted parameters and $X_{\mathrm{obs}}$ is the fitted parameter determined from other observations where $\sigma _{X_{\mathrm{obs}}}$ is the associated error. The multiplicative factor $G(0,1)$ is a Gaussian randomly generated number with a mean of 0 and a standard deviation of 1. This includes in the fit the error determined from prior observations. Equation 1 was extended to include all constraints on all parameters where prior parameter information is known. The procedure is described on a case by case basis for each object in the following sections.          

To find the best-fitting solution a chi-squared minimisation was carried out using the IDL function MPFIT, utilizing the Levenburg-Marquart algorithm. The $1\sigma$ best-fit parameter uncertainties were calculated using a Monte-Carlo method. $10^{5}$ synthetic data sets were created by adding a 1$\sigma$ Gaussian random variable multiplied by the error on the radial velocity to the radial velocity data points. The free parameters were re-optimised for each simulated data-set to obtain the distribution of the best-fit parameter values. The distributions were not assumed to be Gaussian and the 1$\sigma$ limits were found where the distribution enclosed $\pm34.1\%$ of the values away from the median. As a consistency check, the data were also analysed using the RML fitting procedure used by, for example, \citet{Hebrard2011} and \citet{Moutou2011}. 

\section{Analysis}\label{Analyis}

\subsection{WASP-13}

WASP-13b is a sub-Jupiter mass exoplanet with $M_{\mathrm{p}}=0.500 \pm 0.037{M _{\mathrm{J}}}$ and $R_{p}=1.407 \pm 0.052R_{\mathrm{J}}$ with an orbital period of $4.4$ days \citep{Chew2013}. Its host star is a G1V type with $T_\mathrm{eff}=5989 \pm 48$K, $M_{\star}=1.187\pm 0.065 \mathrm{M_{\odot}}$, $\log g=4.16$ and solar metallicity. The host star has a projected rotational velocity of $v \sin i=5.74 \pm 0.38 $ \kms \citep{Chew2013}.
	
\begin{figure*}
\centering
\includegraphics[angle=0, width=\columnwidth]{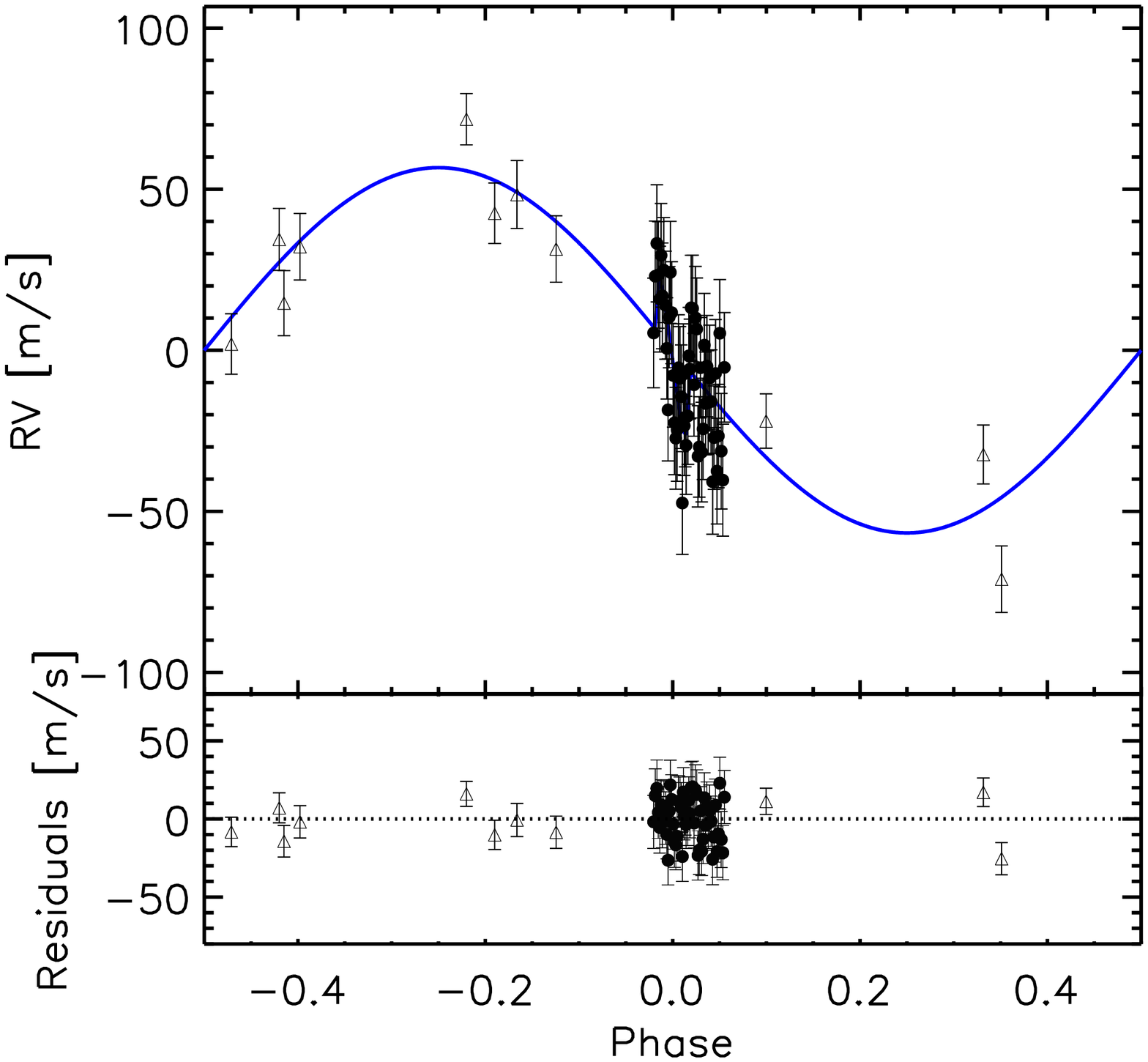}
\includegraphics[angle=0, width=\columnwidth]{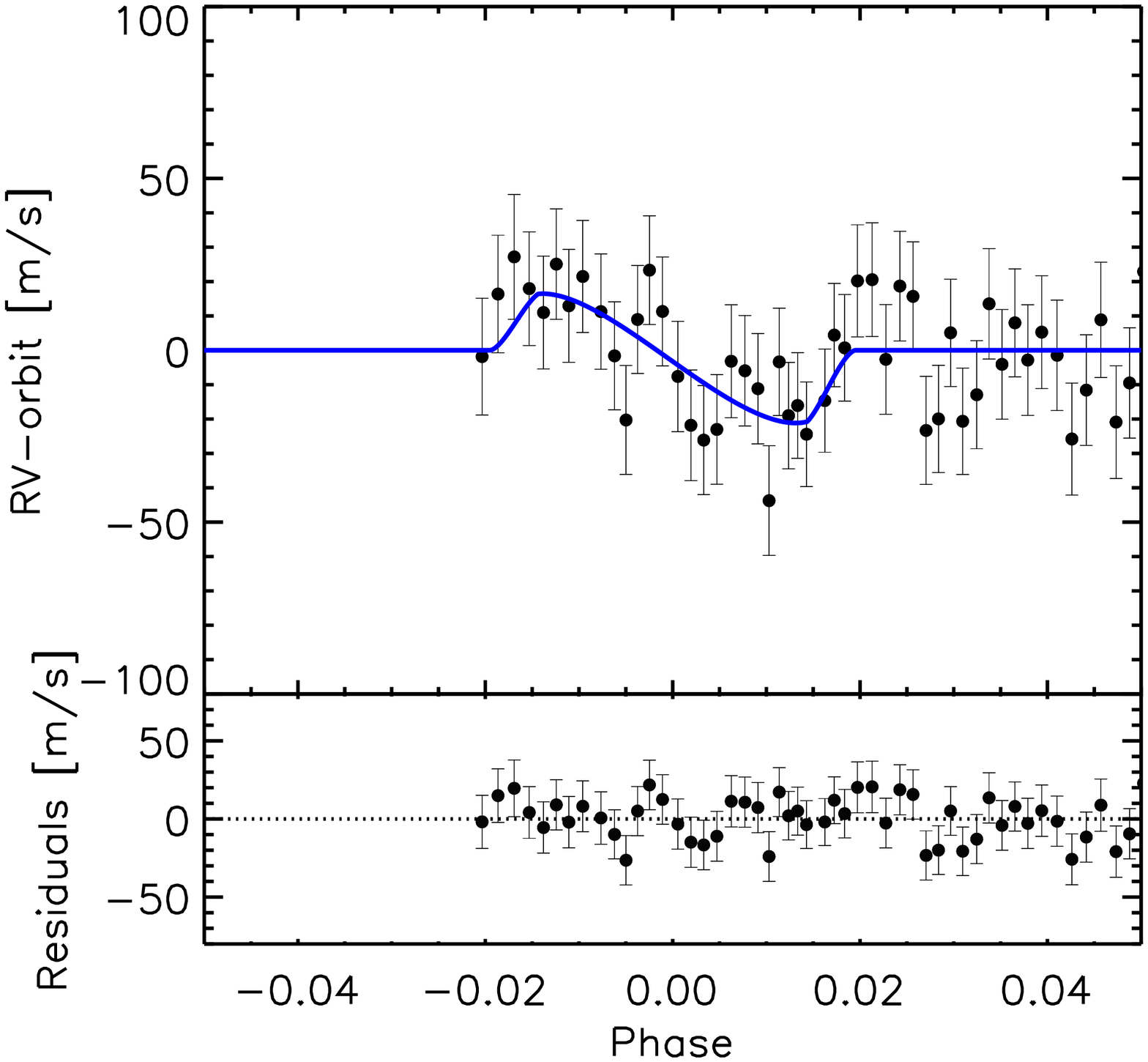}
\vspace{-10pt}
\caption{\emph{Left Panel:} WASP-13 phase-folded orbit minus the systemic velocity overplotted with the best-fitting model (solid line). Out-of-transit radial velocities of \citet{Skillen2009} are displayed as open triangles and the measured radial velocities using SOPHIE at the OHP on the night March 6 2012 are displayed as filled circles. \emph{Right Panel:} Spectroscopic transit minus the orbital velocity, overplotted with the best-fitting RM model with the residuals shown below.}
\label{w13plot}
\end{figure*}

A transit of WASP-13b was observed with the SOPHIE spectrograph at the 1.93-m telescope at the Observatoire de Haute-Provence (OHP) on the night of 2012 March 6. We acquired a total of 54 spectra, 32 spectra in-transit and 22 spectra out-of-transit with a total of 228 minutes of out-of-transit data (primarily post-transit). SOPHIE was used in high efficiency mode (HE) with a resolution of $R=40,000$ and fast read-out mode, maintaining a constant signal to noise throughout the observing run ($S/N=30$). Typical exposure times were 600s and the seeing remained $\sim$2 arcseconds during the observing night.  The measured radial velocities are listed in Table \ref{w13RVs}. Moon illumination was 97\% and at a distance of $\sim$30 degrees from WASP-13 on the night of observation. We note that the lunar RV was 0.002 \kms which, when compared to the systemic velocity of WASP-13 of $\gamma_{\mathrm{orbit}}=9.8345 \pm 0.0031$ \kms, means most of the lunar contribution lies outside the stellar absorption-line profile. Nevertheless, we have applied the standard lunar contamination correction available through the SOPHIE data reduction pipeline. To fit the orbit, we used 11 SOPHIE observations acquired during the discovery of the planet \citep{Skillen2009}. The fitted orbit can be found in the left hand panel of Figure \ref{w13plot} with the fitted systemic velocity $\gamma_{\mathrm{orbit}}=9.8345 \pm 0.0031$ \kms removed from the RV datapoints.

To fit the RM effect the OTS model was used as described in Section 2.2. The fitted RM waveform can be found in the right hand panel of Figure \ref{w13plot} with the systemic velocity $\gamma_{\mathrm{transit}}=9.7854 \pm 0.0037$ \kms removed from the RV datapoints. In the model the linear limb darkening coefficient was chosen from the tables of \citet{Claret2004} (ATLAS models) for the $g'$ filter. A linear interpolation using  John Southworth's JKTLD code with stellar parameters of $T_{\mathrm{eff}}=5989$K, $[M/H]=0.06$, $\log g=4.16$, $v_{\mathrm{mic}}=1.27$ \kms was used and a linear limb-darkening coefficient of $u=0.75$ was adopted. The eccentricity was fixed at $e=0$ and a constraint on $v \sin i=5.74 \pm 0.38$ \kms was added to the $\chi^{2}$ penalty function \citep{Chew2013}.  

A $\chi ^{2}$ statistic was adopted of the form of Equation \ref{eqw1} where the priors included in the penalty function are listed in Table \ref{Adopted}.
\noindent
The fitted parameters and uncertainties are given in Table \ref{Resultsw13}. The best-fitting model is shown in Figure \ref{w13plot} where ${\lambda=8^{\circ}}^{+13}_{-12}$. It is clear from the shape of Figure \ref{w13plot} that the planet has a prograde orbit. In addition, the RV waveform shows a symmetric shape indicating star-planet alignment.
\noindent
A fit was also attempted with no prior on $v \sin i$ with no effect on the fitted parameters. This is explained by the large impact parameter ($b=0.6$) where the degeneracy between $\lambda$ and $v \sin i$ only becomes important in the low impact parameter regime. In this regime the form of the RM signal is not strongly dependent on $\lambda$, however the amplitude is controlled by both $\lambda$ and $v \sin i$. It has been shown that applying a penalty function in this regime has no overall impact on the fitted parameters \citep{Simpson2011a} and this is indeed what we found in the case of WASP-13.

\subsection{WASP-32}

WASP-32 is a massive exoplanet with $M_{p}=3.60 \pm 0.07{\mathrm{M_{J}}}$ and $R_{p}=1.19 \pm 0.06R_{\mathrm{J}}$ in a $P=2.7$ day orbit. The host star has $T_\mathrm{eff}=6100 \pm 100$K, $M_{\star}=1.10 \pm 0.03 \mathrm{M_{\odot}}$, $\log g=4.4$ and is lithium depleted \citep{Maxted2010}. The projected rotational velocity of the host star is $v \sin i=4.8 \pm 0.8$ \kms \citep{Maxted2010}.  

\begin{figure*}
\centering
\includegraphics[angle=0, width=\columnwidth]{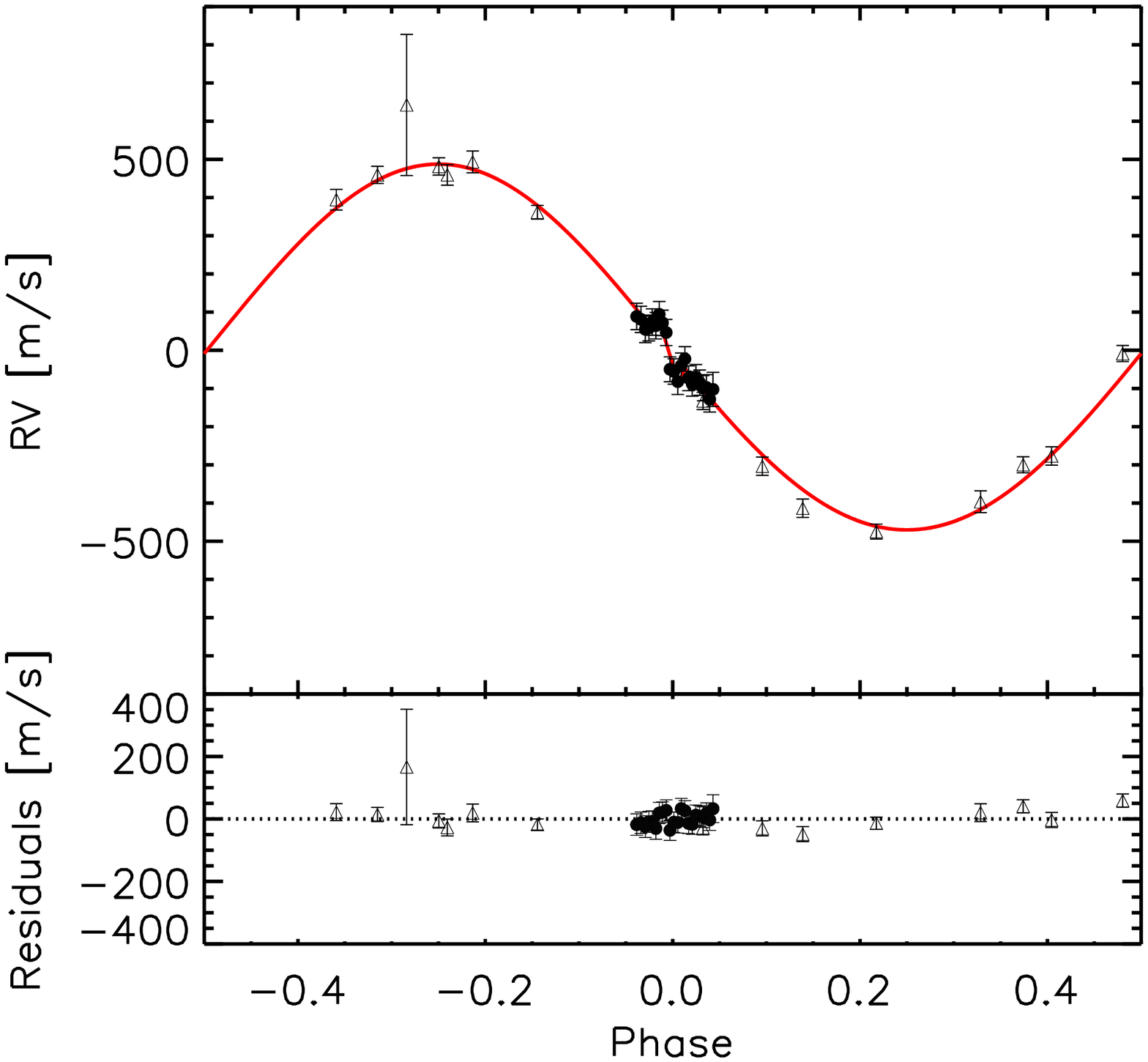}
\includegraphics[angle=0, width=\columnwidth]{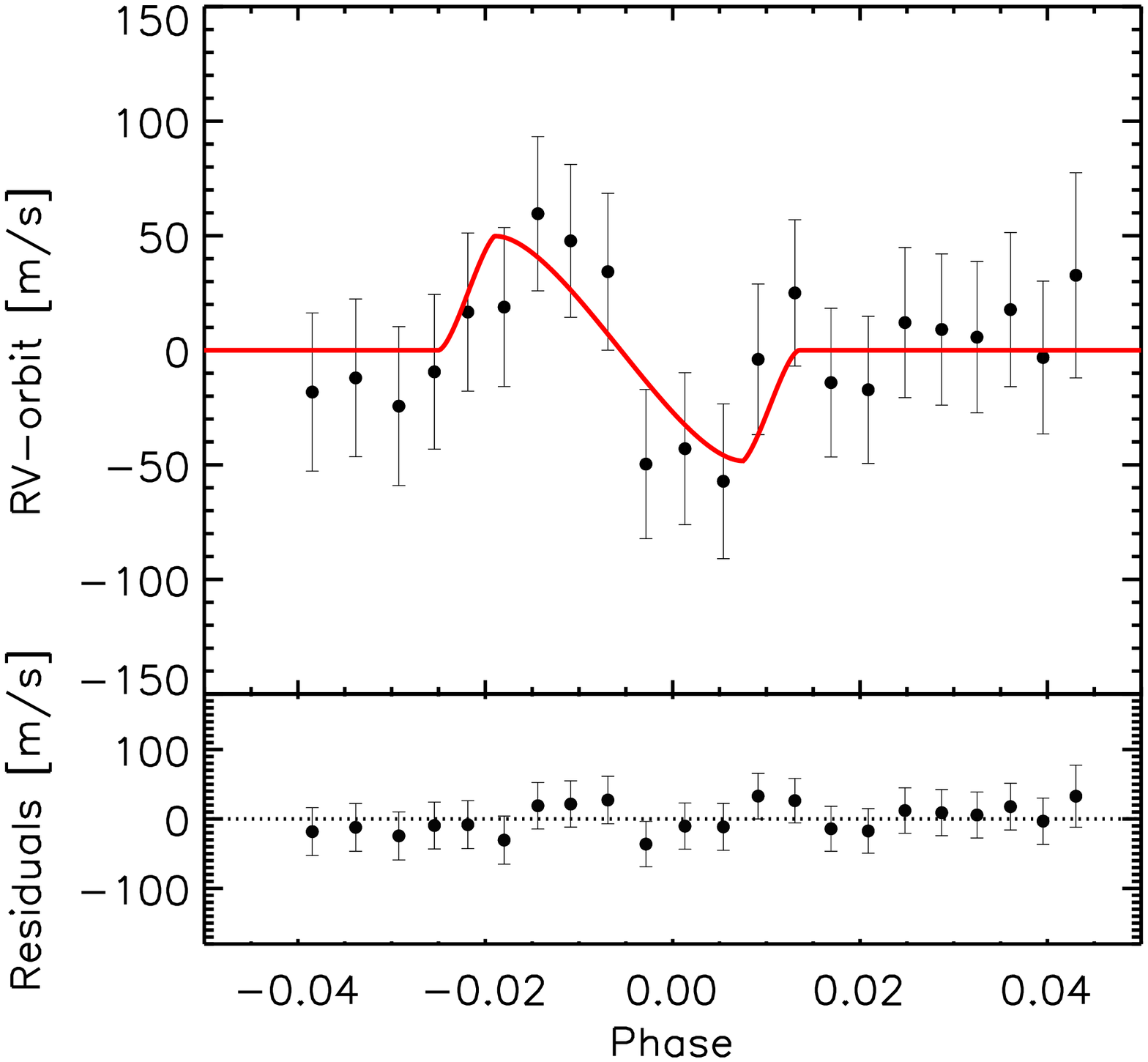}
\vspace{-10pt}
\caption{\emph{Left Panel:} WASP-32 phase-folded orbit minus the systemic velocity overplotted with the best-fitting model (solid line). Out-of-transit radial velocities of \citet{Maxted2010} from CORALIE are displayed as open triangles and the measured radial velocities using SOPHIE are displayed as filled circles. \emph{Right Panel:} Spectroscopic transit minus the orbital velocity, overplotted with the best-fitting RM model with the residuals shown below.}
\label{w32plot}
\end{figure*}

We acquired 22 spectra during the transit of WASP-32, covering the complete transit. 4 spectra were acquired prior to transit and 8 spectra were acquired post transit with a total of 168 mins of observations acquired outside transit. The data were obtained using SOPHIE on the night of 2011 August 29th, with clear conditions and a typical seeing of 2.5 arcseconds. SOPHIE was operated in high-efficiency mode and there was no moonlight pollution on the night of observation. The derived radial velocities can be found in Table \ref{w32RVs}. To fit the orbit 14 CORALIE out-of-transit RVs were used from the WASP-32 discovery paper \citep{Maxted2010}. The fitted orbit can be found in the left hand panel of Figure \ref{w32plot} with the systemic velocity offset $\gamma_{\mathrm{orbit}}=18.2796^{+0.0061}_{-0.0062}$ \kms removed from the RV data points.

To fit the RM effect the OTS model was used as described in Section 2.2. The fitted RM waveform can be found in the right hand panel of Figure \ref{w32plot} with the systemic velocity $\gamma_{\mathrm{transit}}=18.1698 \pm 0.0095$ \kms  removed from the RV data points. We note the difference in the orbital and transit systemic velocities is $\sim$ 100 \ms  for WASP-32 and $\sim$50 \ms for WASP-13, comparable to values obtained for other objects in the literature. In particular, \citet{Simpson2010} measured a difference in the orbital and transit velocities for WASP-1b of $\sim$200 \ms using the same RM model and observational approach. Also we examined the likelihood that the measured systemic velocity offsets could be driven by long-term stellar activity by phase-folding the WASP-13 and WASP-32 light-curves with the transits removed. A clear sinusoidal photometric modulation was detected in the light-curve for WASP-32 at the $\sim$2$\%$ level, although modulations were not detected for WASP-13. The increased activity level of WASP-32 also explains why a period peak was detected in the periodogram (discussed in Section 3.3). Using the relation presented in \citet{Saar1997} the RV shift due to inhomogeneous spot coverage can be estimated. A $\sim$100 \ms RV shift for WASP-32 is expected with $\sim$2$\%$ inhomogeneous spot coverage, comparable to the difference in our reported systemic velocity for WASP-32 compared to the systemic velocity of the orbital data. Therefore, the difference in systemic velocities may be explained by spot coverage. It is important to note that the RM effect duration for WASP-32 is $\sim$2.4 hours, during which the host star rotates by $\sim$8 degrees. Thus, it is unlikely new spot features would rotate into view during transit, and thus systemic velocity offsets over the course of the RM observation are insignificant.     
\noindent
In the model the linear limb darkening coefficient was chosen as before with stellar parameters of $T_{\mathrm{eff}}=6100$K, $[M/H]=-0.13$, $\log g=4.39$, $v_{\mathrm{mic}}=2$ \kms, resulting in a linear limb darkening coefficient of $u=0.71$. WASP-32 is a reasonably eccentric system with $e=0.018 \pm 0.0065$ and this parameter was fixed in the RM fit. 

A $\chi ^{2}$ statistic was adopted of the form of Equation \ref{eqw1}, where the priors included in the penalty function are taken from \citet{Maxted2010} and are listed in Table \ref{Adopted}. The best fitting model is shown in Figure \ref{w32plot} where $\lambda={-2^{\circ}}^{+17}_{-19}$ and the fitted parameters are listed in Table \ref{Resultsw32}. It is clear from Figure \ref{w32plot} that WASP-32 has a symmetric RM waveform, moving from redshift to blueshift, consistent with an aligned prograde orbit. The fitted $\lambda$ is consistent with that found recently by \citet{Brown2012} where $\lambda={10.5^{\circ}}^{+6.4}_{-5.9}$.

The fitted $v \sin i=7.6^{+4.2}_{-3.1}$ \kms is consistent with that found from sprectroscopic fitting, $v \sin i=4.8 \pm 0.8$ \kms \citep{Maxted2010} and the measured $v \sin i=3.9^{+0.4}_{-0.5}$ \kms derived from Doppler Tomography \citep{Brown2012}. However, our determined $v \sin i$ is noticeably larger than the others that have been found. Thus, a fit with a prior on $v \sin i$ set to that found by \citet{Brown2012} was attempted. It was found that $\lambda$ is insensitive to fixing $v \sin i$ in the fit, with little change in $\chi_{\mathrm{red}}^{2}$. Thus, the fit with a prior on $v \sin i$ was taken as our adopted solution with a fitted $\lambda={-2^{\circ}}^{+17}_{-19}$ and $v \sin i=3.9 \pm 0.5$ \kms. Also we attempted a fit using the \citet{Brown2012} HARPS RVs alone and found that the error bars on $\lambda$ were increased relative to the \citet{Brown2012} results. We attribute this to the use of simultaneous photometry in the \citet{Brown2012} analysis but note our fit is the first independent analysis on the alignment of WASP-32b.   

\subsection{3D Alignment Angle}

Modelling the Rossiter-McLaughlin effect leads to a determination of the sky-projected alignment angle, $\lambda$. As a consequence of only measuring the sky-projected alignment angle, in some cases a measured $\lambda$ that indicates an aligned planet may actually be a misaligned system. For example, measuring a $\lambda=0^{\circ}$ does not necessarily indicate an aligned planetary system if the inclination of the stellar rotation axis is unknown. If the stellar rotation axis is inclined relative to the line of sight or if the impact parameter, $b$, is close to 0 then the planet may be misaligned but the symmetry of the RM waveform would indicate an aligned system. A true three dimensional reconstruction of the system geometry can be gleaned if the inclination of the host star can be found simultaneously with the sky-projected angle, removing any possible ambiguities that remain from RM observations alone.

By determining the stellar rotation period ($P_\mathrm{rot}$) combined with the projected rotational velocity ($v \sin i$) and stellar radius ($R_{\star}$), the stellar inclination ($i_{\star}$) can be found via:

\begin{eqnarray} \label{stellarinc}
\sin i_{\star}=P_{\mathrm{rot}} \times \left(\frac{v \sin i_{\star} }{2 \pi R_{\star}}\right) 
\end{eqnarray}

\begin{figure}
\centering
\includegraphics[angle=0, width=\columnwidth]{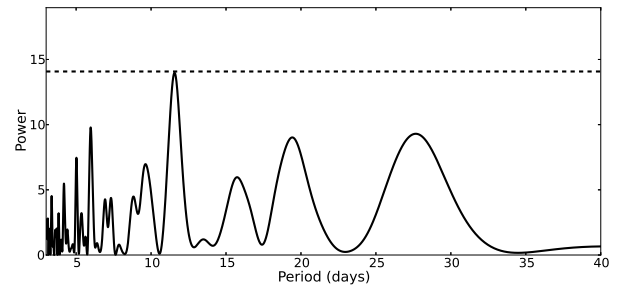}
\vspace{-15pt}
\caption{Lomb-Scargle periodogram analysis for the detrended WASP-32 lightcurve from the SuperWASP data archive, observed in the date range interval $5048.57056713-5153.42813657$ HJD and with planetary transits removed. The dashed line indicates the FAP of $0.1 \%$ and indicates the level where a period detection is defined. The peak period corresponds to $P_\mathrm{rot}=11.6 \pm 1.0$ days and is at the $0.1 \%$ FAP power level. The peak at $5.8$ days is a harmonic of the peak detected period.}
\label{periodwasp32}
\end{figure}

\noindent
The projected stellar rotational velocity, $v \sin i$, and stellar radius estimates can be obtained via spectral analysis. There are a number of ways to determine the stellar rotation period, to determine $i_{\star}$, either by monitoring Ca H and K emission or photometric monitoring of starspots (e.g. \citealt{Simpson2010}, \citealt{ Watson2010}). In our analysis we adopt the latter, where the modification of disk integrated light indicates the passage of starspots across the stellar surface. By sourcing the detrended WASP lightcurves for WASP-13 and WASP-32, an extensive Lomb-Scargle \citep{Lomb1976, Scargle1982} periodogram analysis was carried out to search for significant stellar rotation periods. The significance of the periods was estimated using the false alarm probability (FAP- \citealt{Horne1986}). A detection was defined when the peak in the periodogram surpassed the 0.1\% FAP power level. This means the detected period has a 99.9\% confidence level that it does not arise by chance. Before carrying out the periodogram analysis on all sourced light curves, the updated transit ephemeris was used to remove the planetary transits. This prevented unwanted harmonics entering the periodograms and ensured intrinsic stellar periodicities were analysed. 
\noindent
As a useful comparison, assuming the rotation axis is perpendicular to the line of sight ($i_{\star}=90^{\circ}$), $P_\mathrm{rot}$ can be computed for all three systems using the $v \sin i$ quoted for all three systems in Section 3. In the case of WASP-32 a number of $v \sin i$ measurements have been determined but we use $v \sin i=3.9^{+0.4}_{-0.5}$ \kms from Doppler Tomography \citep{Brown2012} to compute the stellar inclination for WASP-32. For WASP-13 and WASP-32, true alignment would lead to expected values of the stellar rotation period of $P_\mathrm{rot}=17.1$ and $11.7$ days, respectively. No statistically significant rotation period was detected for WASP-13, however a statistically significant period was detected in the WASP-32 data above the 99.9\% confidence level. The detected period is $P_\mathrm{rot}=11.6 \pm 1.0$ days and the periodogram is shown in Figure \ref{periodwasp32}.  With $P_\mathrm{rot}$ known, $i_{\star}$ can be determined from equation 5, where an $i_{\star}=81^{\circ} \pm 9$ was found. This, combined with the planetary inclination ($i_{p}$) determined from the planetary transit and $\lambda$, allows the true 3D alignment angle, $\psi$, to be found via:

\begin{equation}
\cos \psi=\cos i_{\star}\cos i_{p}+\sin i_{\star} \sin i_{p} \cos \lambda  
\end{equation}
\noindent
The measured 3D alignment angle, using $\lambda={10.5^{\circ}}^{+6.4}_{-5.9}$ from \citet{Brown2012} gives $\psi={11^{\circ}} \pm 14$ and using the value of $\lambda={-2^{\circ}}^{+17}_{-19}$, derived in this work, $\psi={2^{\circ}} \pm 16$. Both results indicate that the planet is aligned when considered as a 3D system.

\begin{figure*}
\centering
\includegraphics[angle=0, width=16.5cm,height=13cm, keepaspectratio]{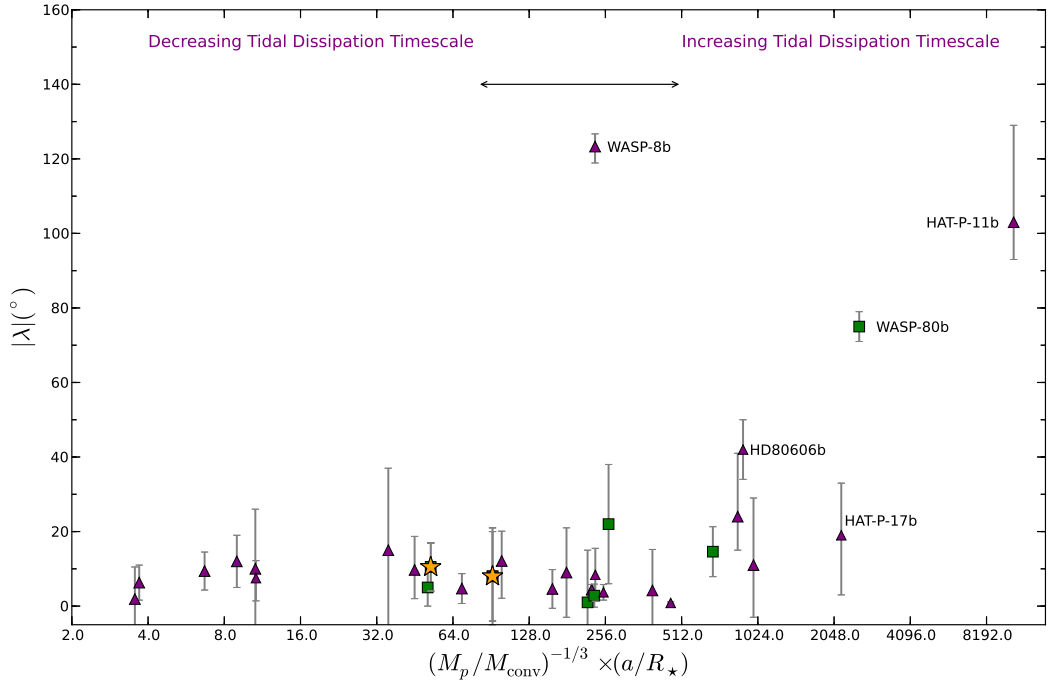}
\vspace{-10pt}
\caption{Plot of $|\lambda|$ against $(M_p/M_{\mathrm{conv}})^{-1/3} \times (a/R_{\star})$ for all systems with $T_{\mathrm{eff}}<6150$K. $a/R_{\star}$ is obtained directly from the planetary transit. The convective mass, $M_{\mathrm{conv}}$, was derived from the EZ-Web stellar evolution code.Systems with age determinations are shown as triangle symbols and those with an assumed age of 4 Gyrs are shown as square symbols. WASP-13b and WASP-32b are shown as starred symbols on the plot.}
\label{tides}
\end{figure*}    

\section{Conclusions} \label{Conc}

The spectrosopic transits of WASP-13b and WASP-32b were observed with the SOPHIE spectrograph and the projected spin-orbit alignment was determined for both systems where $\lambda={8^{\circ}}^{+13}_{-12}$ and ${-2^{\circ}}^{+17}_{-19}$, respectively. WASP-13 and WASP-32 are consistent with alignment within 1$\sigma$. This suggests WASP-13 and WASP-32 had a gentle migration history and remained unperturbed from the original obliquity of the protoplanetary disk. An alternative scenario is the gradual loss of orbital energy by the planet through tidal dissipation, acting to realign the stellar spin and planetary orbital axes over a long enough timescale \citep{Winn2010}. The misalignment angle measured for WASP-32 in this work is consistent with the value of $\lambda={10.5^{\circ}}^{+6.4}_{-6.5}$ measured by \citet{Brown2012}. Our measured 3D alignment angle of $\psi=11^{\circ} \pm 14$ provides further evidence that the system is well-aligned. It is important to note that $\psi$ has not been measured for many systems (see Table \ref{psi}) and WASP-32 adds to the number of systems with a measured 3D alignment angle. Further, Table \ref{psi} shows that some systems are unambiguously aligned. Attempts have been made to derive the original obliquity distribution (Triaud et al., 2010; Li 2013) assuming a $\mathrm{cos} i_{\star}$ probability distribution in the stellar inclination, however this deprojection technique means that cases where $i_{\star}=90^{\circ}$ are unaccounted for. All current measurements of $\psi$ in Table \ref{psi} show that there is a bimodal distribution in $\psi$: a planetary population that is aligned and one that is near isotropic. Thus, any attempt to deproject the population of spin-orbit angles is destined to fail if current trends in $i_{\star}$ are not recognised.    

\citet{Winn2010} proposed one mechanism that could explain the observed distribution of alignment angles via tidal dissipation with the host star. In this scheme aligned planets are expected around cool stars ($T_{\mathrm{eff}}<6250$K) and misaligned planets are expected orbiting hot host stars ($T_{\mathrm{eff}}>6250$K). WASP-13 and WASP-32 have $T_{\mathrm{eff}}=5989 \pm 48$K and $T_{\mathrm{eff}}=6100 \pm 100$K, respectively, and therefore both lie in the cool regime. Thus both WASP-13 and WASP-32 add further evidence to alignment arising from tidal interactions. Also it must be noted that WASP-32 has a $T_{\mathrm{eff}}$ close to 6250K, perhaps indicating that it is possible for massive planets (in this case with a mass $>3 \mathrm{M_{J}}$) to tidally realign around relatively hot host stars.

Alignment is expected to be determined by planet-star tidal interactions. The tidal interaction timescale due to tidal dissipation in the convective envelope is related to $q$, the planet to star mass ratio, and the scaled semi-major axis, $a/R_{\star}$ (see equation 2 of \citealt{Albrecht2013}):

\begin{equation}
\label{tidal}
\frac{1}{\tau_\mathrm{CE}}=\frac{1}{10 \times 10^{9} \mathrm{yr}}q^2\left(\frac{a/R_{\star}}{40}\right)^{-6}
\end{equation}
\noindent
Thus, the above equation shows that $\tau_\mathrm{CE} \propto q^{-2} \times (a/R_{\star})^6$. As planet-star tidal interactions with the convective envelope are thought to be responsible for aligning hot-Jupiters via tidal dampening, we modified Equation \ref{tidal} to include the convective mass of the planet host, $M_{\mathrm{conv}}$. Thus in Figure \ref{tides} an ensemble of systems with RM measurements are plotted against $(M_p/M_{\mathrm{conv}})^{-1/3} \times (a/R_{\star})$, a quantity proportional to the tidal dissipation timescale. The convective envelope mass, $M_{\mathrm{conv}}$, was derived using the EZ-Web stellar evolution code \footnote{EZ-Web Stellar Evolution Code: \newline http://www.astro.wisc.edu/~townsend/static.php?ref=ez-web}. To run the stellar models, the age of the system is required. For systems lacking derived ages, we have assumed an age of 4 Gyrs, but note the results are largely insensitive to age. Note the $x$-axis scale was chosen such that $\tau_\mathrm{CE}^{1/6} \propto (M_p/M_{\mathrm{conv}})^{-1/3} \times (a/R_{\star})$ for plotting convenience. As suggested by \citet{Zahn1977} tides are dissipated more effectively when the planet orbits a star with a convective envelope. \citet{Winn2010} have postulated that tides have changed the distribution of spin-orbit angles with $T_{\mathrm{eff}}<6150$K but left the distribution unaltered with $T_{\mathrm{eff}}>6350$K. Thus, only `cool' systems with $T_{\mathrm{eff}}<6150$K are plotted in Figure \ref{tides}. It can be observed that as the tides become weaker (when the tidal dissipation timescale increases) there is some evidence that misaligned orbits are more likely. WASP-13 and WASP-32 are plotted in Figure \ref{tides} and are consistent with alignment. A recent addition to the ensemble of RM measurements is WASP-80b \citep{Triaud201380}, a K7-M0 star and the coolest host star in the sample with $T_{\mathrm{eff}}=4145 \pm 100$K. Even as the coolest system, the planet is on an inclined circular orbit with $|\lambda|=75^{\circ}$ similar to the spin-orbit angle measured around hotter mid-F stars. This suggests that hot-Jupiters may have been more frequently misaligned in the past. However, other mechanisms could act to misalign a system such as the presence of another peturbing body or if the host star is not old enough to develop a convective envelope. WASP-80b is considered a rare example of a misaligned system around a cool host star \citep{Triaud201380}. However, Figure \ref{tides} suggests that WASP-80b is yet to realign because of its long tidal dissipation timescale. 

Even though the above analysis is simplified, Figure \ref{tides} suggests that planet-star tidal interactions likely play a role in damping the obliquities of hot-Jupiters around cool host stars. Systems with short tidal dissipation timescales are preferentially aligned, however those with longer timescales show an apparent random distribution in $\lambda$. This may suggest that hot-Jupiters once had a broader range of obliquities in the past and, that they have been realigned over time via tidal interactions \citep{Albrecht2013}. In Figure \ref{tides}, WASP-8b is the most obvious outlier in the distribution, however WASP-8 is a dynamically complex system with suggestions the Kozai mechanism or violent dynamical interations may explain the misaligned orbit \citep{Queloz2010}.
      
\begin{table}
\caption{Table of all cases where the 3D alignment angle, $\psi$, has been reported in the literature. The measured $\psi$ and reference is indicated in the table. Multiple references indicate where $\psi$ has been measured in seperate studies. Cases where multiple $\psi$ measurements are listed with a single reference stems from orbital geometry degeneracies.  Our result for WASP-32 adds to the number of systems with a complete 3D alignment angle determination.}	
	\label{psi}
	\begin{tabular}{lcc}
	\hline\hline
	Object			& $\psi~ (^{\circ})$ & Reference\\
	\hline
CoRoT-18b & $20 \pm 20$ & [1]  \\
HAT-P-7 & $>86.1$   & [2] \\
HAT-P-11 & $106^{+15}_{-11},~ 97^{+8}_{-4}$ & [3]     \\
Kepler-16(AB)b &$<18.3$  & [4]  \\
Kepler-17b & $0 \pm 15 $  & [5] \\ 
Kepler-63b & $104^{+9}_{-14}$  & [6]  \\	
Kepler-13.01 & $54 \pm 4,~56 \pm 4,~124 \pm 4,~126 \pm 4$  & [7]   \\
KOI-368.01 & $69^{+9}_{-10}$  & [8]  \\	
PTFO 8-86956b & $73.1 \pm 0.5$  & [9] \\   
WASP-15b & $>90.3$  & [10]   \\  
WASP-17b & $>91.7,~>92.6$	  & [10], [11]  \\
WASP-19b & $<19,~<20$   & [10], [12]  \\   
\textbf{WASP-32b} & $\mathbf{11 \pm 14}$  & \textbf{This Work}  \\
\hline
\hline
\end{tabular}

\medskip
\textbf{[1]} \citet{Hebrard2011psi} \textbf{[2]} \citet{Winn2009} \textbf{[3]} \citet{Ojeda2011} \textbf{[4]}  \citet{Winn2011psi} \textbf{[5]} \citet{Desert2011} \textbf{[6]} \citet{Ojeda2013} \textbf{[7]} \citet{Barnes2011} \textbf{[8]} \citet{Zhou2013} \textbf{[9]} \citet{Barnes2013} \textbf{[10]} \citet{Triaud2010psi}  \textbf{[11]} \citet{Bayliss2010} \textbf{[12]} \citet{Hellier2011}
%\label{psi}
\end{table}

It is known that stars with $\mathrm{M}>1.2\mathrm{M}_{\odot}$ cool as they evolve along the main sequence. As the star cools an outer convective envelope develops, increasing the strength of the tidal interactions. Thus the distribution of spin-orbit angles is expected to change with time where a planet originally on a misaligned orbit will realign as the convective envelope of the host star develops. Triaud 2011 plotted $|\lambda|$ against age for all systems with $\mathrm{M}>1.2\mathrm{M}_{\odot}$. The plot provides weak evidence that the spin-orbit alignment distribution changes with time and is another manifestation of the influence of tidal interactions. Objects with ages 2.5-3 Gyrs appear aligned, whereas more misaligned systems are observed around stars with younger ages. Even though the plots of $|\lambda|$ against $a/R_{\star}$ and age show evidence for evolution due to tides, it is still unclear if an original misaligned hot-Jupiter population would survive realignment around `cool' host stars or tidally infall into the star, leaving the aligned population observed today.          

We have presented RM measurements for WASP-13 and WASP-32. Analysing out-of-transit survey photometry for WASP-32 revealed the rotation period of the host star, and thus the 3D alignment angle $\psi=11^{\circ} \pm 14$ of the planetary system. WASP-32 adds to the number of systems with a full 3D alignment angle determination. It is clear that it is becoming increasingly important to investigate the full star-planet-disk \citep[e.g.,][]{Watson2011, Kennedy2013} alignment in order to fully assess the migration history of exoplanets. Only with an alignment determination of the whole system can we begin to fully evaluate the migration scenarios of hot-Jupiters.       

\section*{Acknowledgments}
We thank J. Southworth for making his JKTLD code available for calculating limb darkening coefficients. This research has made use of the Astrophysics Data System (ADS), the Extrasolar Planets Encyclopaedia, R. Heller's Holt Rossiter-McLaughlin Encyclopaedia and R. Townsend's EZ-Web stellar evolution code. R. D. B. acknowledges support from the Queen's University Belfast Department for Education and Learning (DEL) university scholarship. C.A.W. acknowledges support by STFC grant ST/I001123/1. All RM observations were taken with the SOPHIE spectrograph on the 1.93 m telescope at Observatoire de Haute-Provence (CNRS), France. WASP-13 was observed as part of directors discretionary time and WASP-32 as part of OPTICON time. R. D. B. acknowledges the complementary analysis of WASP-13 and WASP-32 RM observations by \'{E}lodie H\'{e}brard using an RML fitting routine. This provided a consistency check of our determined obliquities. Also R. D. B. would like to thank D. R. Anderson for simultaneously fitting the photometry and RVs for all our targets, providing a further consistency check. R. D. B. would like to thank Armaury Triaud for useful advice that greatly enhanced the discussion in this report. We would also like to thank the anonymous referee for their careful reading and detailed comments on the paper.

\bibliographystyle{mn2e}
\bibliography{brothwell_2013bib}

   \begin{table*} 
	\centering
	\caption{Radial velocities and 1$\sigma$ error bars of WASP-13 measured with \textit{SOPHIE} during and outside transit.}
	
	\begin{tabular}{ccc}
	\hline \hline
	BJD		&RV		&Error	\\
	-2 400 000	&(\kms)	&(\kms)	\\
	\hline
5993.31409  &    9.7904 &    0.0145 \\
5993.32150  &    9.8080 &    0.0146 \\
5993.32902  &    9.8182 &    0.0155 \\
5993.33602  &    9.8084 &    0.0141 \\
5993.34263  &    9.8009 &    0.0140 \\
5993.34859  &    9.8145 &    0.0137 \\ 
5993.35443  &    9.8019 &    0.0140 \\
5993.36088  &    9.8099 &    0.0139 \\
5993.36932  &    9.7990 &    0.0143 \\
5993.37564  &    9.7856 &    0.0134 \\
5993.38098  &    9.7665 &    0.0135 \\
5993.38634  &    9.7953 &    0.0134 \\ 
5993.39186  &    9.8092 &    0.0135 \\
5993.39793  &    9.7967 &    0.0135 \\
5993.40506  &    9.7772 &    0.0137 \\
5993.41112  &    9.7625 &    0.0137 \\ 
5993.41708  &    9.7577 &    0.0135 \\
5993.42317  &    9.7603 &    0.0136 \\
5993.42985  &    9.7796 &    0.0140 \\
5993.43624  &    9.7763 &    0.0137 \\
5993.44225  &    9.7706 &    0.0137 \\
5993.44745  &    9.7376 &    0.0136 \\
5993.45216  &    9.7776 &    0.0133 \\
5993.45649  &    9.7616 &    0.0132 \\
5993.46069  &    9.7642 &    0.0131 \\
5993.46483  &    9.7555 &    0.0130 \\
5993.47332  &    9.7646 &    0.0128 \\
5993.47773  &    9.7833 &    0.0128 \\
5993.48256  &    9.7792 &    0.0132 \\
5993.48852  &    9.7982 &    0.0139 \\
5993.49531  &    9.7980 &    0.0141 \\
5993.50170  &    9.7743 &    0.0136 \\
5993.50820  &    9.7951 &    0.0136 \\ 
5993.51431  &    9.7916 &    0.0135 \\
5993.52041  &    9.7521 &    0.0134 \\
5993.52615  &    9.7550 &    0.0133 \\
5993.53177  &    9.7796 &    0.0133 \\
5993.53739  &    9.7534 &    0.0132 \\
5993.54392  &    9.7606 &    0.0134 \\
5993.54966  &    9.7866 &    0.0137 \\
5993.55566  &    9.7685 &    0.0136 \\
5993.56168  &    9.7801 &    0.0134 \\
5993.56763  &    9.7688 &    0.0137 \\
5993.57412  &    9.7764 &    0.0140 \\
5993.58127  &    9.7691 &    0.0137 \\
5993.58813  &    9.7442 &    0.0139 \\
5993.59483  &    9.7579 &    0.0137 \\
5993.60164  &    9.7778 &    0.0143 \\
5993.60865  &    9.7475 &    0.0140 \\
5993.61494  &    9.7584 &    0.0137 \\
5993.62145  &    9.7903 &    0.0142 \\
5993.62827  &    9.7537 &    0.0153 \\
5993.63558  &    9.7447 &    0.0148 \\
5993.64269  &    9.7797 &    0.0145 \\
	\hline
	\end{tabular}
	\label{w13RVs}
	\end{table*}

   \begin{table*} 
	\centering
	\caption{Radial velocities and 1$\sigma$ error bars of WASP-32 measured with \textit{SOPHIE} during and outside transit.}
	\begin{tabular}{ccc}
	\hline \hline
	BJD		&RV		&Error	\\
	-2 400 000	&(\kms)	&(\kms)	\\
	\hline
5803.44375 &    18.2586 &    0.0230 \\
5803.45634 &    18.2510 &    0.0229 \\
5803.46888 &    18.2249 &    0.0231 \\
5803.47919 &    18.2285 &    0.0225 \\
5803.48888 &    18.2439 &    0.0230 \\
5803.49942 &    18.2344 &    0.0231 \\
5803.50922 &    18.2643 &    0.0224 \\
5803.51874 &    18.2419 &    0.0222 \\
5803.52952 &    18.2165 &    0.0228 \\
5803.54057 &    18.1203 &    0.0217 \\
5803.55190 &    18.1145 &    0.0221 \\
5803.56303 &    18.0880 &    0.0225 \\
5803.57313 &    18.1301 &    0.0219 \\
5803.58378 &    18.1474 &    0.0213 \\
5803.59428 &    18.0968 &    0.0216 \\
5803.60508 &    18.0819 &    0.0214 \\
5803.61574 &    18.0997 &    0.0218 \\
5803.62640 &    18.0852 &    0.0220 \\
5803.63663 &    18.0709 &    0.0220 \\
5803.64635 &    18.0726 &    0.0224 \\
5803.65582 &    18.0417 &    0.0222 \\ 
5803.66534 &    18.0676 &    0.0298 \\
	\hline
	\end{tabular}
	\label{w32RVs}
	\end{table*}

\newpage
\begin{table*}
	\caption{Derived system parameters and uncertainties for WASP-13. The effective temperature is taken from \citet{Chew2013}. Fitted free parameters are listed with the corresponding errors followed by the parameters controlled by priors (listed in Equation 2).}	
	\label{Resultsw13}
	\begin{tabular}{lc r@{}l}
	\hline\hline
	Parameter (units)			&Symbol & \multicolumn{2}{c}{Value}\\
	\hline
	Free parameters: & & \\[4pt]
	Projected alignment angle (\degr)				&$\lambda$		&\,8	&$^{+13}_{-12}$	\\[4pt]
	Projected stellar rotation velocity (\kms)	& $v\sin{i}$ 		&\,5.7	& $\pm$0.4    		\\[4pt]
	RV semi-amplitude 	(\kms)					& $K$ 			&\,0.0564 &$\pm$0.0043 \\ [4pt]
	Systemic velocity of SOPHIE transit dataset (\kms)	& $\gamma_{\mathrm{transit}}$		&\,9.7854		&$\pm$0.0037 \\[4pt]
	Systemic velocity of SOPHIE orbital dataset (\kms)	& $\gamma_{\mathrm{orbit}}$ 	&\,9.8345		&$\pm$0.0031 \\[4pt]
	\hline
	Parameters controlled by priors: & &\\[4pt]
	Period (days) 						&$P$		&\,4.3530135	&$\pm$0.000003	\\[4pt]
	Transit epoch	($\mathrm{BJD_{UTC}}$ - 2 400 000) 		&$T_\mathrm{0}$	&\, 5304.53998	&$\pm$0.00025\\[4pt]
	Planet/Star radius ratio				& $R_{p}/R_{*}$	&\,0.0918&$^{+0.0127}_{-0.0126}$ \\[4pt]
	Scaled semi-major axis				&$a/R_{*}$ 		&\,7.54&$\pm$0.27	\\[4pt]
	Orbital inclination (\degr)			& $i$  &\,	85.43	&$\pm$0.29		\\[4pt]
	\hline
	Fixed parameters: & & \\[4pt]
	Eccentricity 				&$e$				& $0\,$ 				&\,				\\[4pt]
	Limb darkening 			&$u$			&$0.75\,$				&\,						 \\
	\hline
	Effective temperature (K)				    & $T_{\mathrm{eff}}$				& 5989\, & $\pm$48		\\
	\hline
	\vspace{0.2cm}
	\end{tabular}
	\end{table*}
	
	\begin{table*}
	\caption{Derived system parameters and uncertainties for WASP-32. The effective temperature is taken from Maxted et al. (2010)}	
	\label{Resultsw32}
	\begin{tabular}{lc r@{}l}
	\hline\hline
	Parameter (units)			&Symbol & \multicolumn{2}{c}{Value}\\
	\hline
	Free parameters: & & \\[4pt]
	Projected alignment angle (\degr)				&$\lambda$		&\,-2			&$^{+17}_{-19}$	\\[4pt]
	RV semi-amplitude 	(\kms)					& $K$ 			&\,0.4789  &$^{+0.0079}_{-0.0078}$		\\[4pt]
	Systemic velocity of SOPHIE transit dataset (\kms)	& $\gamma_{\mathrm{transit}}$	&\,18.1698		&$\pm 0.0095$		\\[4pt]
	Systemic velocity of SOPHIE orbital dataset (\kms)	& $\gamma_{\mathrm{orbit}}$ 	&\,18.2796	&$^{+0.0061}_{-0.0062}$	\\[4pt]
	\hline
	Parameters controlled by priors: & &\\[4pt]

Projected stellar rotation velocity (\kms)			& $v\sin{i}$ 		&3.9\,		&$\pm 0.5$     		\\[4pt]
	Period (days) 						           &$P$		 &\,2.7186590	 &$\pm 0.000008$		\\[4pt]
	Transit epoch	(HJD - 2 400 000)   &$T_\mathrm{0}$	  &\,5150.39051 &$\pm 0.00050$ \\[4pt]
	Planet/Star radius ratio				& $R_{p}/R_{*}$	&\,0.1091 &$\pm 0.0010$		\\[4pt]
	Scaled semi-major axis				&$a/R_{*}$ 		&\,7.63		& $\pm 0.35$ 		\\[4pt]
	Orbital inclination (\degr)				& $i$ 			&\,85.30		&$\pm 0.50$		\\[4pt]
	\hline
	Fixed parameters: & & \\[4pt]
	Eccentricity 				&$e$				& $0.018\,$ 				&$\pm 0.0065$\, \\[4pt]
	Limb darkening 			&$u$				&$0.71\,$				&\,						 \\
	\hline
	Effective temperature (K)						& $T_{\mathrm{eff}}$				& 6100\, & $\pm$ 100		\\
	\hline
	\end{tabular}
	\end{table*}	
	
\end{document}